\documentclass{article}
\usepackage{graphicx}

\usepackage[dcucite]{hharvard}

\begin{document}

\title{Eprints and the Open Archives Initiative\footnote{Published 
as Library Hi Tech, Volume 21, Number 2, 151-158 (2003), 
DOI: 10.1108/07378830310479794.
Journal version 2003/03/14, this version: $ $Date: 2003/07/04 03:24:35 $ $}}

\author{Simeon Warner\\
Computing and Information Science, Cornell University, NY, USA.\\
{\tt simeon@cs.cornell.edu}}
\date{ }
\maketitle

\begin{abstract}
The Open Archives Initiative (OAI) was created as a practical way to 
promote interoperability between eprint repositories. Although the scope 
of the OAI has been broadened, eprint repositories still represent a 
significant fraction of OAI data providers. In this article I present
a brief survey of OAI eprint repositories, and
of services using metadata harvested from eprint repositories using
the OAI protocol for metadata harvesting (OAI-PMH).
I then discuss several situations where metadata harvesting may be used
to further improve the utility of eprint archives as a component of 
the scholarly communication infrastructure.
\end{abstract}

{\bf Keywords:}
Eprint, Open Archives Initiative, Metadata Harvesting, Scholarly Communication

\section{Introduction}

The Open Archives Initiative (OAI) was born from the 1999 Santa Fe 
Universal Preprint Service meeting~\hcite{SantaFeUPSMeeting}
and the Santa Fe Convention~\hcite{SantaFeConvention}, with the 
intention of improving scholarly communication through improved
interoperability between eprint archives. During the
first year of discussion and development the scope of the OAI was
extended and the associated protocol generalized to be more widely 
applicable~\hcite{Lagoze+2002}. The current, application-neutral, 
Open Archives Initiative Protocol for Metadata Harvesting 
(OAI-PMH)~\hcite{OAI-PMHv2.0} is the result of almost three years of
experimentation and development. 

While the focus of the OAI has broadened to include more than just eprints, the
original participants have continued to play active roles in the development
of the OAI. One result is that there are a growing number of eprint archives
for which metadata is available via the OAI-PMH.

A recent study~\hcite{DLFOutsell2002} of the scholarly information environment
reported that 87\% of respondents (faculty, graduate and undergraduate students) 
use online methods to find print materials
for research (spread between online databases, library finding aids, search
engines, subject directories and other forms). This suggests that the inclusion
of OAI-PMH harvested metadata for eprints in library finding aids, search engines 
and subject directories would increase the impact of eprints -- with the added 
advantage for users that they can also access the material online. The same 
study reported that the 74\% of respondents said that online access 
was their preferred access method for electronic journal articles.
(As opposed to using a library terminal, or personal holdings.)

\section{What is an eprint?}

Different writers use the terms {\it eprint}\/ in more or less general senses. 
Some imply a very general meaning, for example: 
``An Eprint Archive is a collection of digital documents''~\cite{EprintsOrgFAQ},
which would include a private digital library or a proprietary electronic
journal with restricted access.
Others restrict the term to author-self archived electronic documents, 
and yet others apply the term only to author-self archived pre-prints.
The term was originally used in the announcement of ``Algebraic Geometry
E-Prints'' at Duke. Paul Ginsparg recounts
``...originally `e-print' was a pun on preprint, originally appeared on a page
created by Dave Morrison at Duke in Feb 1992 for `Algebraic Geometry E-Prints',
the second archive based on my original hep-th csh scripts. (on that page,
Dave credited his colleague Greg Lawler with coining the word.)
the word `e-print' then quickly devolved to meaninglessness but more recently
has been rehabilitated to mean an article either in draft or final form
{\it self-archived} by the author.''~\cite{eprint-origin}.

In this paper I use a definition similar to that given by
Pinfield {\it et al}\/~\hcite{Pinfield+2002}. 
I use the term {\it eprint}\/ to group together many 
forms of scholarly literature for which there is open access to 
the full-content via the internet. Eprints may include: journal 
articles, pre-prints, technical reports, books, theses and dissertations.
Eprints may or may not be refereed.

\section{Who uses eprints?}

The importance of eprints varies widely over different subject areas. 
Eprints have been most successful in high-energy physics where they are
used as a dissemination mechanism that shortcuts the delays and 
access-restrictions associated with conventional journals.
This success is usually attributed to the pre-existing 
culture of sharing pre-prints~\hcite{Kreitz+1996,Kling+2000}.
Kling and McKim~\hcite{Kling+2000} point out that other 
disciplines use electronic media in different ways and for different parts 
of the scientific communication process. They suggest that many of the differences 
between disciplines will be ``durable features of the scholarly landscape''.
We should thus avoid the temptation to imagine that simply extending
the model of physicists' use of eprints to other disciplines will succeed.
However, the OAI framework is not dependent on one model of electronic
media use and can provide a discipline independent infrastructure for 
metadata exchange while also supporting the exchange of discipline specific
metadata. 

\section{Review and certification}

Publication and review are not the same thing in
general, and one does not necessarily imply the other.
To publish means simply ``to make generally known'' or 
``to disseminate to the public''~\cite{Websters}. If we consider 
the four components of scholarly communication described by 
Roosendaal and Guerts~\hcite{Roosendaal+1998} ---
Registration, Certification, Awareness and Archiving --- 
registration is the component satisfied by publication, and 
certification may be satisfied by peer-review or some other
process (perhaps more than one).

The arXiv eprint archive~\cite{arXiv} provides almost no certification
and yet has transformed scholarly communication in some areas of
physics. arXiv does provide the registration, awareness and, to
some extent, the archiving components of scholarly communication. 
However, even those physicists that rely on arXiv for dissemination
of their research usually also rely on conventional journals to
provide the certification (by peer-review) necessary to support
career advancement and funding applications.

Theses and dissertations are different from typical research 
articles in that they are subject to a certification process
that is usually quite separate from publication and any associated
revenue stream. This makes theses and dissertations one form of
scholarly communication that already appear as eprints over
a much broader range of subjects than are covered by the few
successful discipline-based eprint repositories. The
meaning of the certification of a theses or dissertation depends 
strongly on the reputation of the degree awarding institution so 
any metadata which will allow a user to assess the certification
should include this information. The draft metadata format proposed
by the Networked Digital Library of Theses and Dissertations 
(NDLTD)~\hcite{NDLTD-Metadata} includes specific fields to 
describe the name, level, discipline and grantor of the degree.

One hurdle for more widespread acceptance of eprints
is users' skepticism about the authenticity and credibility of 
information obtained from the Internet. A recent Digital Library 
Federation (DLF) commissioned 
study~\hcite{DLFOutsell2002} reported that more than half of respondents
say they verify the accuracy of information they obtain from the Internet.
The same report shows a wide spectrum of methods used to determine
the authoritativeness of information obtained from the Internet: 
19\% only reference known sources, 14\% check with alternative
sources, 13\% trust the author, 9\% trust the sponsoring organization or
publisher, 9\% trust the web site, 7\% only reference academic sources
provided by an accredited institution. These figures suggest that the
identity of an eprint repository and the authority associated with it
will be important in determining acceptance.

There already exists a metadata format specifically for including  
``branding'' information with OAI-PMH records~\hcite{OAI-PMH-Branding}. 
The branding information includes an icon and a link to be associated 
with the icon (typically the home page of the originating repository). The linked icon 
can be displayed by service providers which use harvested metadata.
Greater adoption of this standard may help support the 
projection of repository identities and associated authority assumptions
in OAI based services. This may be particularly effective for institutional
repositories where the institution carries significant authority.

\section{Eprints and the OAI}

The recent SPARC white paper on institutional repositories~\hcite{SPARC-IR-WP}
presents a compelling case for institutional repositories, a type of eprint 
archive, as part of an evolving scholarly publishing system. Interoperability
as provided by the existing OAI infrastructure is cited as an 
essential infrastructure component required for the effective use of
such repositories. 

Perhaps the most prominent system to encourage the creation of eprint
repositories is the EPrints software~\cite{EprintsSoftware}, 
the goal of which is to promote 
author self-archiving and institutional archives. Exchange of metadata 
via the OAI-PMH is a key element of the EPrints software and has been 
built into the system since its 
first version. At the very least, by sharing metadata, individual 
repositories will be included in the `union catalogs' upon which OAI search 
and alerting services are based. In this way, institutional repositories keep 
the publication and preservation functions with the institution, 
which has motivations to support these activities, while still allowing 
the eprints to be part of a global collection. The development of 
additional OAI-based services can add value to the entire collection
or to selected segments. 

Table~\ref{OAIDPTable} shows that over half of the registered 
OAI data providers (repositories) contain metadata about eprints. 
Determination of whether a repository contains metadata about
eprints was made based on the repository name, description and
response to the OAI-PMH Identify verb. Only repositories 
registered with the OAI website were included. 
Some repositories are registered twice because they support both 
versions 1.1 and~2.0 of the OAI-PMH. These duplicate entries and 
``test servers'' were excluded from the percentages calculated.

\begin{table}[htb]
\begin{center}
\begin{tabular}{|l|r|r|}
\hline
Data provider type              & number & \% of total \\
\hline
Metadata about eprints          & 57     & 54\% \\
Metadata not about eprints      & 30     & 29\% \\
Unreachable or broken           & 18     & 17\% \\
Duplicate v1.1 and v2.0 servers & 9      & not included \\
Test servers                    & 5      & not included \\
\hline
\end{tabular}
\end{center}
\caption{\label{OAIDPTable} Survey of registered OAI data providers
to estimate number that contain metadata about eprints (October 2002)}
\end{table}

Table~\ref{OAIDataTable} shows the number of items (equivalently, the 
number of metadata records in the mandatory Dublin Core format) in a 
sample of OAI repositories and an estimate of the 
number which describe eprints.
The arXiv.org~\cite{arXiv} eprint archive holds the largest collection 
of metadata about eprints (and the full content of those eprints)
currently exported via the OAI-PMH. 
The OCLC xtcat repository~\cite{xtcatOAI} has many more records than 
arXiv.org, over 4 million, but exceedingly few include any means by 
which the full content may be accessed. 
Of the OAI data providers that export metadata for eprints, a significant
fraction are repositories of theses and dissertations. Many of these
repositories have just a few hundred records at present. 
The VTETD repository is a repository of electronic theses and 
dissertations with a significant number of records (3665) and 
approximately two-thirds include links to freely accessible 
full content.

\begin{table}[htb]
\begin{center}
\begin{tabular}{|l|r|r|l|}
\hline
Data provider               &  Items   &  Eprints & Resource type \\
\hline
arXiv~\cite{arXivOAI}       &  212,976 &  212,976 & articles / tech. reports / theses \\
NCSTRLH~\cite{NCSTRLHOAI}   &    20517 &    20517 & articles / tech. reports \\
NACA~\cite{NACAOAI}         &     7549 &     7549 & tech. reports \\
RePEc~\cite{RePECOAI}       &  231,822 & $\sim$6000$^1$ & articles / tech. reports/ \\
                            &          &          & author \& institution records \\
LTRS~\cite{LTRSOAI}         &     3002 &     3002 & articles / tech. reports \\
VTETD~\cite{VTETDOAI}       &     3665 &  2408$^2$& theses \\
CogPrints~\cite{CogPrintsOAI}&    1543 &     1543 & articles \\
BioMed Central~\cite{BMCOAI}&     1186 &     1186 & articles \\
CULEuclid~\cite{EuclidOAI}  &     4938 &      114 & articles \\
\hline
\end{tabular}
\end{center}
{\small
$^1$ Estimate provided by Christian Zimmermann after discussion on the
repec-run mailing list \\
{\tt http://lists.openlib.org/pipermail/repec-run/2002-November/000557.html} \\
$^2$ The metadata records are marked with `restricted', `unrestricted' or
`mixed'. This number is the count of metadata records marked `unrestricted'. \\
$^3$ The count of eprints includes only those articles which will remain 
`open access', a significant additional number of articles are currently
listed as `limited time open access'. 
}

\caption{\label{OAIDataTable} Survey of selected OAI data providers with 
significant fraction of metadata about eprints (October 2002)}
\end{table}

Table~\ref{OAIServicesTable} is a survey of all registered OAI service
providers. The list is dominated by search services, some of which have
additional facilities. For example, torii~\cite{torii} includes 
personalization and personal document storage facilities.

\begin{table}
\begin{center}
\begin{tabular}{|l|l|l|}
\hline
Service provider       & Coverage           & Service \\
\hline
arc~\cite{arc}         & OAI repositories   & search \\
my.OAI~\cite{myOAI}    & 11 OAI DPs         & search with personalization \\
Perseus~\cite{Perseus} & A few OAI repositories + & search (full text for local) \\
                       & local & \\
OAIster~\cite{OAIster} & OAI repositories$^1$ & search \\
iCite~\cite{iCite}     & arXiv$^2$          & search by author and citations \\
torii~\cite{torii}     & OAI eprints$^3$    & search, personalization, and \\
                       &                    & personal document store \\ 
PKP~\cite{PKP}         & OAI eprints$^4$    & search \\
citebaseSearch~\cite{Citebase} & A few repositories$^5$ & search with local citation \\
                       &                    & and impact analyses \\
Scirus~\cite{Scirus}   & A few OAI repositories + & search \\ 
                       & proprietary + web & \\
NCSTRL~\cite{NCSTRL}   & A few OAI repositories + & search \\
                       & local$^6$  & \\
\hline
\end{tabular}
\end{center}
{\small 
$^1$ Harvests from sources where full content is available digitally: the 
resources ``have a corresponding web-based digital representation 
(e.g., this would not include the metadata records for slides when the slides 
cannot be accessed through the web).'' \\
$^2$ Limited to the physics section of arXiv (contains most of the 
submissions in arXiv). Harvests full-content outside of OAI-PMH. \\
$^3$ arXiv, JHEP (not currently a registered OAI data provider), BioMed Central, 
M2DB (a local multi-media database). \\
$^4$ Not yet operational, just a few records harvested. \\
$^5$ Test service, harvests full-content PDF from arXiv, BioMed Central and 
CogPrints outside of the OAI-PMH for reference extraction. \\
$^6$ Harvests from institutional CS technical-report repositories 
and includes the locally stored, historical NCSTRL collection.
}
\caption{\label{OAIServicesTable} Survey of end-user services provided by 
registered OAI service providers (October 2002)}
\end{table}

\section{Metadata harvesting as a way to improve the utility of eprints}

At present there are just a few significant eprint archives and 
they tend to be discipline based. This means that researchers in a 
particular field can know about the one or two archives appropriate 
to their interests, and services using metadata from eprint archives can 
manually select appropriate archives to harvest from. If eprint 
archives become more numerous then these conditions will no longer hold.

The OAI and, more recently, the Budapest Open Access Initiative~\hcite{BOAI}
has spurred growth in the number of institutional archives. Presumably some
institutional archives will combine electronic theses and dissertations with
research articles. There are currently more electronic theses and dissertations
archives than general institutional archives registered as OAI data providers.
This is perhaps to be expected because efforts to encourage the use of
electronic theses and dissertations have been going for longer. These
efforts also have the advantage that universities usually have considerably
more control over students  than over faculty, some even mandate the 
electronic submission of theses and dissertations. It is likely that 
universities will have to put significant effort into promoting general
institutional archives and assisting faculty in using them.

The OAI-PMH is designed to support automation and this feature will 
become more important as the number of OAI data providers increases.  
In the next few sections I highlight a few areas where the OAI metadata
harvesting infrastructure may improve the utility of eprints and
eprint archives.

\subsection{Discovery}

There are already several search engines based all or in-part on OAI harvested  
metadata. Some of these services also include local or web data and clearly one
can see adding appropriate OAI metadata to a local search engine as a way of
adding value to the local search and helping to make that a good starting point
for information discovery.

Metadata based search engines are good at answering certain types of question 
which make use of the structure of the metadata. Trivial examples are
``find documents authored by Fred Bloggs'' (notwithstanding problems created by
possible lack of name-authority information to associate articles with author listed
as ``F Bloggs'' and to separate from articles authored by some other Fred Bloggs), or
``find documents that cite A'' (given appropriate citation metadata). 
However, questions such as ``find documents similar to document X'' are 
unlikely to be answered well by query engines which do
not have access to the full content or more complete summary information.
The OAI has steered away from specifying facilities for the exchange of
full content for various reasons, including: appropriate use and 
rights issues, concerns about resource and bandwidth 
use, and the desire to create a strong base-line interoperability 
framework with as many players as possible. One way to 
improve discovery tools without going as far as sharing full content 
is to include summary metadata. Research
articles in many fields typically include author-created abstracts, 
in other cases it may be appropriate to augment metadata with
automatically generated summaries as surrogates for
the full content. 

\subsection{Grouping and classification}

Resources from different sources and in different subject areas are 
often classified according to different classification schemes.
Within an interoperable framework there will be the need to provide 
classifications across disparate collections and to group or classify 
documents according to criteria that were not imagined by the creators
of the source repositories. In most cases it will not be feasible to 
use human catalogers. 

In recent years there have been significant advances in systems 
automatic classification and clustering of 
text-documents~\hcite{Sebastiani2002}. While these
systems cannot yet compete with well-trained human catalogers, they
do produce useful classifications which can be inexpensively applied to
large collections.
A large class of automated classifiers and clustering algorithms are 
based on analysis of word frequencies in the abstracts or full texts 
of documents. An obvious way to permit such analysis is to allow 
full content to be harvested and indexed but this
has at least two potential problems: 1) the content may
be prohibitively large and harvesting it may place excessive burdens on
the resources of data providers, and 2) if full content is harvested, the
harvester must understand every format available to be able to extract the
text. 
Generation of summaries or text-only versions of the content (without markup)
may prove an effective alternative which can both reduce bandwidth
requirements and use local knowledge of content and formats to do these
extraction jobs well.

\subsection{Reference and citation linking}

Citebase~\cite{Citebase} is an impressive demonstration of a reference
and citation linking service built with automatically extracted
reference data from eprints.
While this data can be obtained only from the full content (PDF in the
case of Citebase) of articles, the reference data may then be shared.
All reference and citation data extracted by Citebase is available
via the OAI-PMH, it may be harvested and used by other services.
Citebase also provides a search service which can rank results by
an impact measure based on citations and web usage data.
 
There is the exciting potential to realize a globally connected 
web of eprints based on link extraction and identifier resolution.
Metadata exchanged via OAI-PMH will be one component of such a
web and may be combined with context-sensitive linking services
based on the OpenURL standard~\hcite{OpenURL}.

\subsection{Rights management}

Rights management is important both to commercial entities who
wish to protect their resources, and also to providers of eprints and other 
open-access content who may still be wish to describe appropriate uses
or restrictions on use. Project RoMEO~\cite{RoMEO} is specifically
investigating the rights issues surrounding open-access scholarly 
literature in the context of the OAI. 

Current eprint repositories often include free-text rights statements 
regarding the use of their metadata and content. The creation of tools
which understand rights information requires the development of
machine readable rights metadata. Creation and inclusion of rights metadata
was been suggested several times during the development but rejected
as outside the scope of the core protocol and likely to be difficult or
impossible given the broad scope of OAI. This task may be tractable 
within the limited scope of the eprints community. 

\subsection{Preservation}

The preservation of paper copies of journals and other scholarly 
works has traditionally been a role of university and national 
libraries. The situation is much less clear with commercial 
electronic journals and even worse for eprints. Empirical studies 
of the persistence of objects in digital libraries~\hcite{Nelson+2001} 
and of web references in the scientific literature~\hcite{Lawrence+2001} 
have reported losses of $\sim 3$\%. Such losses would clearly be 
unacceptable in a traditional library. These studies remind us 
that many valuable digital resources, including scholarly works, 
are not adequately provided with preservation strategies.

The LOCKSS~\hcite{LOCKSS} system illustrates one suggested approach 
for the preservation of web published material. 
Using LOCKSS, libraries act to preserve material they
subscribe to by running persistent caches. In the event that material is not
available from the publisher, multiple copies will remain in these 
persistent caches. In the subscription world there is a need for
agreements between publishes and libraries to grant the libraries the
right to make and use cached copies in this way. With open-access 
material, including eprints, there need not be agreements between
individual data providers and agents preserving the content provided
there is some machine readable information that says such caching
is acceptable. In this way, a portal could automatically cache all
material that its users access and the OAI-PMH would be one way to exchange 
additional metadata to support this. Indeed, such a system must be
automatic if it is to work with perhaps thousands of individual 
repositories.

The Open Archival Information System (OAIS)~\hcite{OAIS} provides a 
comprehensive preservation metadata framework that is likely to be
too heavyweight to be implemented by eprint repositories in the near
future. However, various elements described in this framework could 
be used to fulfill specific preservation requirements.
For example, an MD5 checksum~\hcite{MD5} might be included in the
metadata for a resource to allow the resource's validity to be 
checked (part of the fixity requirement in OAIS).  

\section{Conclusion}

There are an increasing number of eprint repositories, both discipline
based and institutional. The Open Archives Initiative (OAI) is a key
infrastructure component that avoids individual repositories becoming
isolated islands of information, by supporting the creation
user services as a separate layer. A number of services already provide 
cross-repository searching and other facilities based on OAI
harvested metadata.

There is considerable potential to add value to eprint repositories
with new services and facilities. Some of these services will require
additional metadata, and this metadata must be machine-readable so that 
it can used automatically. As the number of OAI repositories and
services increases, it will become increasingly important that 
services do not need to create and implement a custom solution for 
each repository they cover.
There is the realizable potential to create a global network of 
interoperating eprint archives with a rich set of discovery,
classification and linking services based on the OAI framework.

\section{Acknowledgments}

This work is supported by the U.S. National Science Foundation under 
agreement number 0132355 and grant number IIS-9817416.

\renewcommand{\refname}{Notes}

\end{document}